\documentclass[fleqn,12pt,twoside]{article}
\usepackage{espcrc1}

\usepackage{graphicx}
\usepackage[figuresright]{rotating}

\newcommand{\km}{$K^-$}

\newcommand{\pim}{$\pi^-$}

\newcommand{\kmpim}{($K^-$,$\pi^-$)}

\newcommand{\pipkp}{($\pi^+$,$K^+$)}

\newcommand{\Lilam}{$^7_\Lambda$Li}

\newcommand{\Belam}{$^9_\Lambda$Be}

\newcommand{\Cthirteenlam}{$^{13}_\Lambda$C}
\newcommand{\Blam}{$^{11}_\Lambda$B}

\newcommand{\Btenlam}{$^{10}_\Lambda$B}
\newcommand{\Nlam}{$^{15}_{\Lambda}$N}
\newcommand{\Olam}{$^{16}_{\Lambda}$O}

\title{
$\gamma$-Ray Spectroscopy in $\Lambda$ Hypernuclei
}

\author{
H.~Tamura,\thanks{Supported by Grant-In-Aid for Scientific Research from
Ministry of Education of Japan, No.11440070 and 15204014.}$^{1}$
~S.~Ajimura,$^{2}$ 
~H.~Akikawa,$^{3,}$\footnote{\uppercase{P}resent address:
\uppercase{J}apan \uppercase{A}tomic 
\uppercase{E}nergy \uppercase{R}esearch 
\uppercase{I}nstitute, \uppercase{T}okai 319-1195, \uppercase{J}apan.}  
~D.E.~Alburger,$^{4}$ 
~K.~Aoki,$^{5}$ 
~A.~Banu,$^{6}$ 
~R.E.~Chrien,$^{4}$ 
~G.B.~Franklin,$^{7}$ 
~J.~Franz,$^{8}$  
~Y.~Fujii,$^{1}$ 
~Y.~Fukao,$^{3}$
~T.~Fukuda,$^{9}$ 
~O.~Hashimoto,$^{1}$ 
~T.~Hayakawa,$^{2}$
~E.~Hiyama,$^{5}$
~H.~Hotchi,$^{9,\dag}$  
~K.~Imai,$^{3}$
~W.~Imoto,$^{9}$ 
~Y.~Kakiguchi,$^{5}$ 
~M.~Kameoka,$^{1}$ 
~T.~Kishimoto,$^{2}$ 
~A.~Krutenkova,$^{10}$
~T.~Maruta,$^{5}$
~A.~Matsumura,$^{1}$ 
~M.~May,$^{4}$  
~S.~Minami,$^{2}$
~Y.~Miura,$^{1}$ 
~K.~Miwa,$^{3}$ 
~T.~Miyoshi,$^{1,}$\footnote{\uppercase{P}resent address: 
\uppercase{U}niversity of \uppercase{H}ouston, \uppercase{H}ouston, 
\uppercase{TX} 77204-5506, \uppercase{USA}.}  
~K.~Mizunuma,$^{1}$ 
~T.~Nagae,$^{5}$
~S.N.~Nakamura,$^{1}$ 
~K.~Nakazawa,$^{11}$ 
~M.~Niiyama,$^{3}$ 
~H.~Nomura,$^{1}$ 
~H.~Noumi,$^{5}$
~Y.~Okayasu,$^{1}$  
~S.~Ota,$^{3}$ 
~T.~Ohtaki,$^{9}$ 
~H.~Outa,$^{5,}$\footnote{\uppercase{P}resent address: 
\uppercase{RIKEN}, \uppercase{W}ako 351-0198, \uppercase{J}apan.} 
~P.~Pile,$^{4}$
~B.P.~Quinn,$^{7}$ 
~A.~Rusek,$^{4}$ 
~P.K.~Saha,$^{9}$
~Y.~Sato,$^{5}$ 
~T.~Saitoh,$^{6}$ 
~M.~Sekimoto,$^{5}$ 
~R.~Sutter,$^{4}$
~H.~Takahashi,$^{3,}$\footnote{\uppercase{P}resent address: 
\uppercase{KEK}, \uppercase{T}sukuba 305-0801, \uppercase{J}apan.}
~T.~Takahashi,$^{1}$ 
~L.~Tang,$^{12}$  
~K.~Tanida,$^{13}$
~S.~Terashima,$^{3}$ 
~M.~Togawa,$^{3}$
~A.~Toyoda,$^{5}$
~M.~Ukai,$^{1}$ 
~H.~Yamauchi,$^{1}$
~L.~Yuan,$^{12}$ 
~S.H.~Zhou$^{14}$\\
\vspace{0.8em}
$^1$~Department of Physics, Tohoku University, Sendai 980-8578, Japan\\
$^2$~Department of Physics, Osaka University, Toyonaka 560-0043, Japan\\
$^3$~Department of Physics, Kyoto University, Kyoto 606-8502, Japan\\
$^4$~Brookhaven National Laboratory, NY 11973, USA\\
$^5$~Institute of Particle and Nuclear Studies, KEK,
Tsukuba 305-0801, Japan\\
$^6$~GSI, Darmstadt D-64291, Germany\\
$^7$~Carnegie Mellon University, Pittsburgh, PA 15213, USA\\
$^8$~Department of Physics, University of Freiburg, Freiburg 79104, Germany\\
$^9$~Osaka Electro-Communication University,
Neyagawa, 572-8530 Japan\\
$^{10}$~Institute for Theoretical and Experimental Physics, Moscow, 
117218 Russia\\
$^{11}$~Department of Physics, Gifu University, Gifu 501-1193, Japan\\
$^{12}$~Department of Physics, Hampton University, Hampton, VA 23668, USA\\
$^{13}$~RIKEN, Wako 351-0198, Japan\\
$^{14}$~China Institute of Atomic Energy,
Beijing 102413, China\\
~~~~(E930('01), E509, E518 collaborations)
}

\begin{document}

\maketitle

\begin{abstract}
The present status of hypernuclear $\gamma$-ray spectroscopy 
with Hyperball is summarized.
We observed two $\gamma$ transitions 
of {\Olam}(1$^- \to 1^-,0^-$) and obtained the strength of the
$\Lambda$$N$ tensor force. 
In $^{10}$B($K^-,\pi^-\gamma$) data,
we did not observe the spin-flip M1 transition of
{\Btenlam}($2^- \to 1^-$), but
$\gamma$ rays from hyperfragments such as
{\Lilam}($7/2^+ \to 5/2^+$) 
and {\Belam}($3/2^+ \to 1/2^+$) were observed.
In $^{11}$B($\pi^+,K^+\gamma$) data,
we observed six $\gamma$ transitions of {\Blam}.
We also attempted an inclusive $\gamma$-ray 
measurement with stopped $K^-$ beam.
\end{abstract}

\section{Introduction}

Since 1998, our project of 
hypernuclear $\gamma$ spectroscopy with 
a germanium (Ge) detector array, Hyperball,
has brought great progress in hypernuclear physics
by revealing precise structure of several light $\Lambda$ hypernuclei
with a resolution in the keV range. 

Hyperball is a large-acceptance germanium (Ge) detector array
dedicated to hypernuclear $\gamma$ spectroscopy.
It has a large efficiency of 2.5 \% at 1 MeV
realized with fourteen large-volume Ge detectors,
and is featured by special readout electronics which 
enables detection of $\gamma$ rays under extremely  
high counting-rate conditions in hypernuclear experiments
with meson beams. Details of Hyperball are 
described in Ref.~\cite{TAM00,TANthesis}.

One of the
the most important physics motivations of hypernuclear
$\gamma$ spectroscopy 
is the study the $\Lambda$$N$ interaction.
In particular, we have been investigating
the strengths of the $\Lambda$$N$ spin-dependent 
forces from precise level structure
of $p$-shell $\Lambda$ hypernuclei.
The potential of the $\Lambda$$N$ two-body effective interaction 
can be written as:

\begin{eqnarray*}
V_{\Lambda N}^{eff}(r) = V_0(r) + 
V_\sigma(r)  \mbox{\boldmath $\sigma$}_\Lambda \mbox{\boldmath $\sigma$}_N 
+ 
V_\Lambda(r) {\bf l}_{\Lambda N} \mbox{\boldmath $\sigma$}_\Lambda 
+ V_N(r)  {\bf l}_{\Lambda N} \mbox{\boldmath $\sigma$}_N   
+ V_T(r) \left[ 3 (\mbox{\boldmath $\sigma$}_\Lambda {\bf
  \hat{r}})(\mbox{\boldmath $\sigma$}_N {\bf \hat{r}}) 
  - \mbox{\boldmath $\sigma$}_\Lambda \mbox{\boldmath $\sigma$}_N \right] 
\label{equation}
\end{eqnarray*}
The four spin-dependent terms, 
namely, the spin-spin term $V_\sigma$, the $\Lambda$-spin-dependent 
spin-orbit term  $V_\Lambda$,
the nucleon-spin-dependent spin-orbit term $V_N$, and the tensor term
$V_T$, have not been studied well by 
experiments of the {\kmpim} and {\pipkp} reaction spectroscopy.  
The radial integrals 
of $V_\sigma$, $V_\Lambda$, $V_N$, and $V_T$
with the $p_N$$s_\Lambda$ wavefunction in $p$-shell hypernuclei
are denoted as $\Delta$, $S_\Lambda$, $S_N$, and $T$, respectively.
These effective-interaction parameters can be experimentally determined 
from low-lying level energies of $p$-shell hypernuclei \cite{DAL78,MIL85}.
However, because of a small level spacing between the spin-doublet states,
high-resolution $\gamma$-ray spectroscopy with Ge detectors 
is necessary to investigate them.

\subsection{ Recent Experiments }

\begin{table}[b]
\caption{List of experiments for hypernuclear $\gamma$ 
spectroscopy with Hyperball since 1998.
\label{EXPLIST}
}
\begin{tabular}{llllll}
\hline
\hline
Experiment & \hspace*{-1mm}Year & \hspace*{-2mm}Line & \hspace*{-2mm}Target/Reaction & Hypernuclei studied  \\
\hline
\hspace*{-2mm}KEK E419 & 1998 & K6 & \hspace*{-1mm}$^7$Li($\pi^+,K^+\gamma$) & {\Lilam} 
 & \hspace*{-2mm}\cite{TAM00,TAN01,SAS04}\\
\hspace*{-2mm}BNL E930('98) & 1998 & D6 & \hspace*{-1mm}$^9$Be($K^-,\pi^-\gamma$) & {\Belam} 
 & \hspace*{-2mm}\cite{AKI02}\\
\hspace*{-2mm}BNL E930('01) & 2001 & D6 & \hspace*{-1mm}$^{16}$O($K^-,\pi^-\gamma$) & {\Olam}, {\Nlam} 
 &  \hspace*{-2mm}\cite{UKA03}\\
 &    &  & \hspace*{-1mm}$^{10}$B($K^-,\pi^-\gamma$) & 
{\Btenlam}, {\Belam}, {\Lilam} etc.&  \\
\hspace*{-2mm}KEK E509 & 2002 & K5 & 
\hspace*{-1mm}$^7$Li,$^9$Be,$^{10}$B,$^{11}$B,$^{12}$C($K^-_{stop},\gamma$) 
& hyperfragments ({\Lilam})  & \hspace*{-2mm}\cite{TAN03}  \\
\hspace*{-2mm}KEK E518 & 2002 & K6 & \hspace*{-1mm}$^{11}$B($\pi^+,K^+\gamma$) & {\Blam}
 &  \hspace*{-2mm}\cite{MIU03}\\
\hline
\hline
\end{tabular}
\end{table}

Table \ref{EXPLIST} shows all the Hyperball 
experiments we have carried out.
In 1998, we performed two experiments,
KEK E419 for {\Lilam} \cite{TAM00,TAN01,SAS04}
and BNL E930('98) for {\Belam} \cite{AKI02},
as we already reported in the HYP2000 conference.
After that, we carried out the second run of BNL E930
(E930('01)) for $^{16}$O and $^{10}$B targets.
With the $^{16}$O target data, 
we observed {\Olam} and {\Nlam} $\gamma$ rays as described later.
With the $^{10}$B target data, 
we did not observe the {\Btenlam}($2^- \to 1^-$) 
$\gamma$ ray transition, but observed several $\gamma$ rays 
from hyperfragments such as {\Lilam} and {\Belam}.
In 2002, we moved Hyperball from BNL to KEK and performed 
two experiments, E509 for hyperfragments \cite{TAN03}
and E518 for {\Blam} \cite{MIU03}.

\begin{figure}[t]
\begin{center}
\includegraphics[height=11cm]{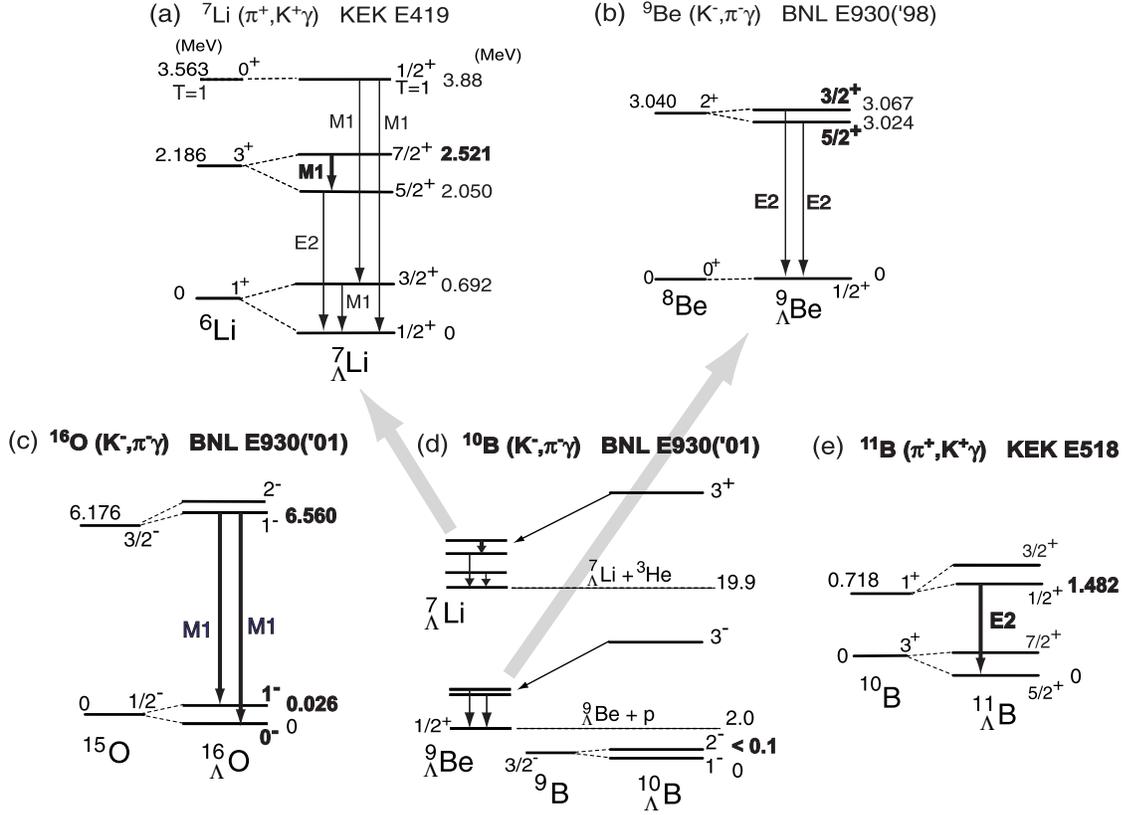}
\vspace*{-0.7cm}
\caption{Level schemes of {\Lilam}, {\Belam}, {\Btenlam}, {\Blam},
and {\Olam} determined from Hyperball experiments. 
Newly observed $\gamma$ rays, measured level energies,
and assigned spins in the recent experiments (E930('01), E509, E518)
are shown in thick arrows and bold letters.
\label{ALLLEVEL}
}
\end{center}
\end{figure}

Figure \ref{ALLLEVEL} shows the level schemes of
$p$-shell hypernuclei determined from these Hyperball experiments.
The $\gamma$ rays first observed and identified in E930('01), E509,
and E518, which are shown in thick arrows,
are described in detail in the following sections.

\section{ {\Belam} and the spin-orbit force}

The experiment BNL E930 aims at determination of
all the spin-dependent
force strengths from structure of several $p$-shell hypernuclei.
Using high-intensity and pure $K^-$ beam at 0.93 GeV/$c$ 
provided by the D6 beam line at BNL AGS, hypernuclei were produced by the
{\kmpim} reaction.
The momenta of incident {\km} and scattered {\pim} were measured with 
magnetic spectrometers to 
obtain the hypernuclear excitation spectrum.
$\gamma$ rays were detected with Hyperball installed around the
target.

\begin{figure}[t]
\begin{center}
\includegraphics[height=4.5cm]{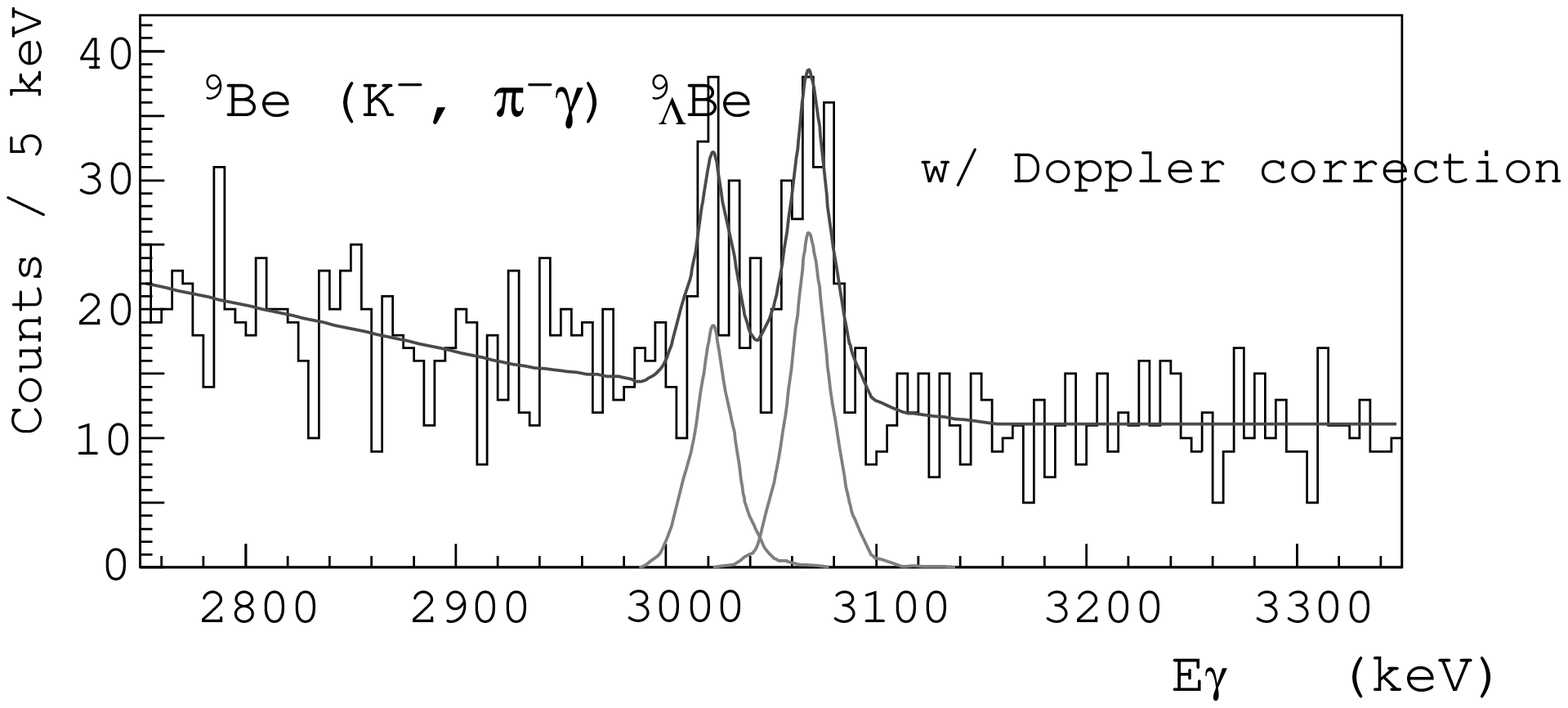}
\vspace*{-1cm}
\caption{ Doppler-shift corrected spectrum of
{\Belam} $\gamma$ rays around 3 MeV
obtained in the E930('98) experiment.
The two-peak structure was well fitted by the simulated peak shape
with a lifetime of $<0.1$ ps. 
\label{BeDOP}
}
\vspace{0.7cm}
\includegraphics[height=4.5cm]{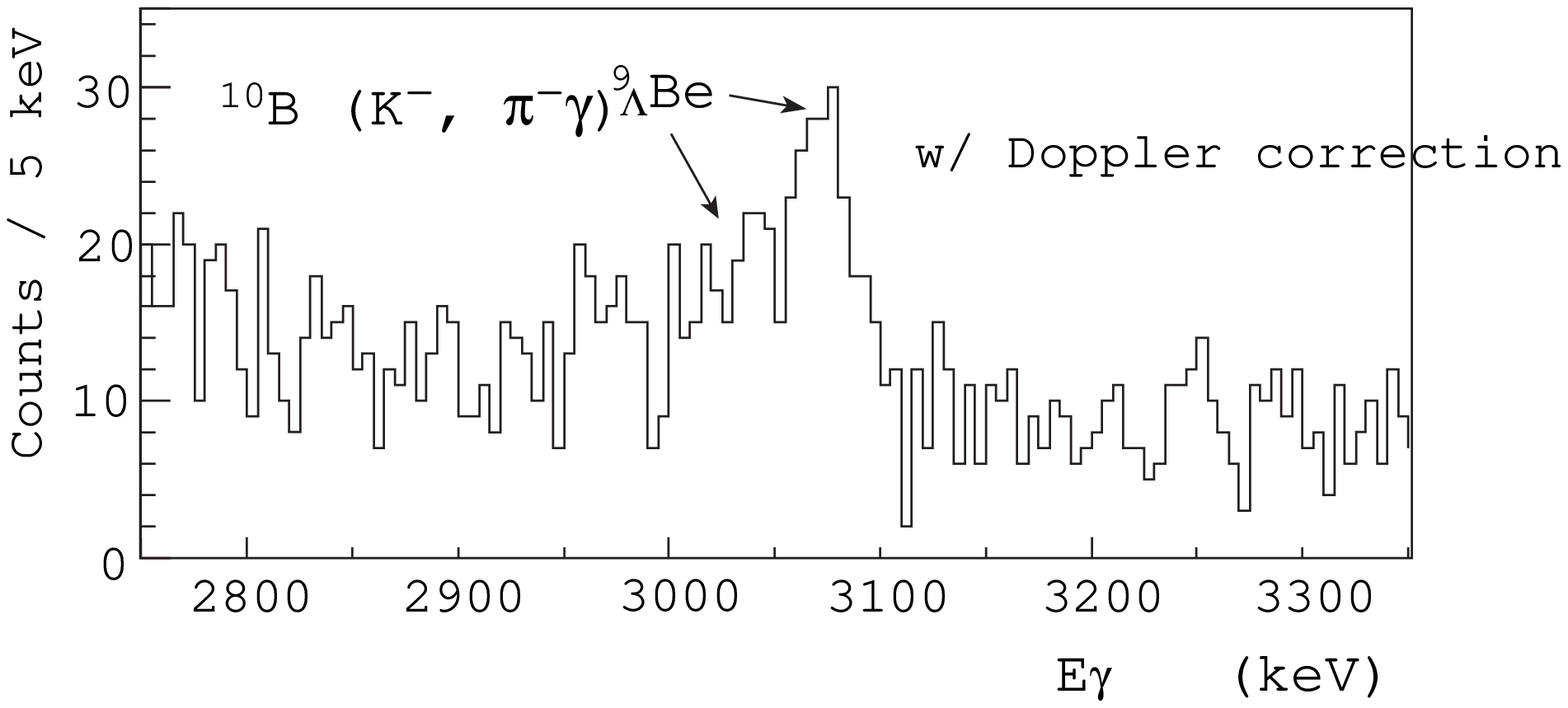}
\vspace*{-0.5cm}
\caption{ $\gamma$-ray spectrum for the mass region 
slightly higher than the bound-state region of {\Btenlam}
($-18 < -B_\Lambda < 28$ MeV).
A $\gamma$-ray peak from {\Belam} is observed.
\label{BE_FRAGMENT}
}
\vspace*{-0.4cm}
\end{center}
\end{figure}

We previouly reported   
that the $^9$Be target data in E930('98) 
exibited the
$5/2^+,3/2^+ \to 1/2^+$ transitions
and revealed
a hypernuclear fine structure of {\Belam}($5/2^+,3/2^+$)
\cite{AKI02}.
Recently, we have applied Doppler-shift correction to this {\Belam}
spectrum and observed clearly-separated two peaks
as shown in Fig.~\ref{BeDOP}.
This structure was well fitted by the simulated peak shape
with a short lifetime ($< 0.1$ ps). The $\gamma$-ray energies 
were obtained to be $3024 \pm 3 \pm 1$ and $3067 \pm 3 \pm 1$ keV,
and the separation energy to be $43 \pm 5$ keV.
The separation energy
and the lifetime have been revised from 
the previous values in Ref.~\cite{AKI02}.

In the $^{10}$B target data taken in E930('01),
we observed a $\gamma$-ray peak at 3065 keV,
when the mass region slightly higher than the {\Btenlam} 
bound states is selected, as shown in Fig.~\ref{BE_FRAGMENT}.
{\Belam} can be produced by proton emission from
excited states of {\Btenlam},
and this observed $\gamma$-ray
energy coincides with the energy of one of the 
{\Belam} $\gamma$ rays observed in the $^9$Be target run.
Therefore, this $\gamma$ ray
is assigned as one of the {\Belam} transitions of
$3/2^+ \to 1/2^+$ or $5/2^+ \to 1/2^+$.
The {\Btenlam}($3^-, -B_\Lambda\sim 1$ MeV) state, which is
expected to have a large cross section,
mostly decays into {\Belam}($3/2^+$) + $p$ (see Fig.~\ref{ALLLEVEL} (d)), 
while other {\Btenlam} excited states decaying into
{\Belam}($5/2^+$) + $p$ have much smaller cross sections \cite{MIL04}. 
Therefore, we assigned the observed peak as
$3/2^+ \to 1/2^+$.
From this spin assignment, the previous result 
for the $\Lambda$-spin-dependent spin-orbit force parameter
of $-0.02 < S_\Lambda < 0.03$ MeV \cite{AKI02} was improved to
$ -0.02< S_\Lambda < -0.01$ MeV.
This sign of the $\Lambda$-spin-dependent spin-orbit term
is consistent with the
$p_{1/2}-p_{3/2}$ spin-orbit splitting in {\Cthirteenlam} 
measured with NaI counter arrays by the BNL E929 
experiment~\cite{AJI01}.

\section{ {\Olam} and $\Lambda$$N$ tensor force}

The purpose of the $^{16}$O target run in E930('01)
is to investigate the $\Lambda$$N$ tensor force strength ($T$),
which has never been studied experimentally.
Since the one-pion exchange is forbidden in the 
$\Lambda$$N$ interaction,
the tensor force is expected to be small, but 
the kaon exchange
and the two-pion exchange through the $\Sigma$-$\Lambda$ coupling
are expected to give some contribution to the
tensor force. 

It was pointed out that
energy spacings of the spin doublets in $p_{1/2}$-shell hypernuclei
are sensitive to the $\Lambda$$N$ tensor force strength \cite{DAL78}.
According to a shell-model calculation by Millener,
the spacing of the ground-state doublet ($0^-$,$1^-$) 
of {\Olam} is given as~\cite{MIL04}
\begin{eqnarray}
E(1^-)-E(0^-) = -0.382\Delta+1.378S_\Lambda-0.004S_N+7.850T+\Lambda\Sigma 
{\rm ~(MeV)},
\label{Eq:16O}
\end{eqnarray}
where $\Lambda\Sigma$ denotes the effect of $\Lambda-\Sigma$ coupling. 
By the $^{16}$O{\kmpim} reaction, we can populate 
the 6 MeV-excited {\Olam} [$(p_{3/2})_n^{-1}(s_{1/2})_\Lambda]_{1^-}$ state 
and detect $M1$ transitions from this state 
to each member of the ground-state doublet,
even if the spacing is too small ($< 100$ keV)
to detect the spin-flip $M1$ transition between the doublet members 
(see Fig.~\ref{ALLLEVEL} (c)).
In addition, the 11 MeV-excited $[(p_{1/2})^{-1}_n (p_{1/2})_\Lambda]_{0^+}$ 
state and
the 17 MeV-excited $[(p_{3/2})^{-1}_n (p_{3/2})_\Lambda]_{0^+}$ 
state of {\Olam} are expected to decay
to excited states of {\Nlam} by proton emission
with sizable branching ratios, which is followed by
emission of {\Nlam} $\gamma$ rays.
The ground-state doublet spacing of {\Nlam},
which also has a large contribution of the $\Lambda$$N$ tensor force, 
may also be measured.

The experimental method and setup are almost
identical to those in the previous E930 run for {\Belam}
described in Ref.~\cite{AKI02}.
We used a 20 cm-thick water target and
irradiated it with 4.0$\times$10$^{10}$ $K^-$ in total.
More description on this experiment is found in Ref.~\cite{UKA03}.

\begin{figure}[t]
\begin{minipage}{10.5cm}
\hspace*{-3mm}\includegraphics[width=10.3cm]{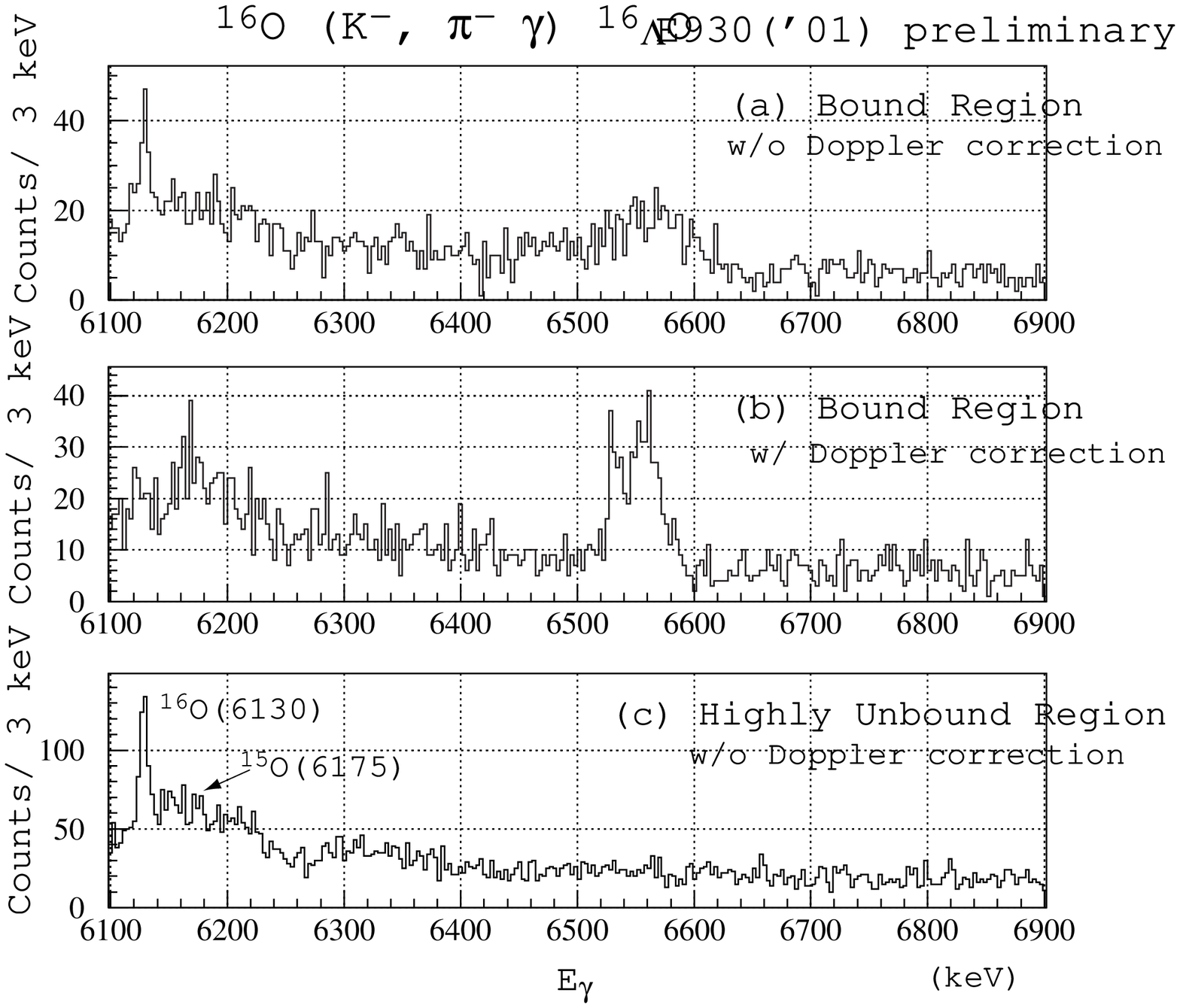}
\vspace{-0.8cm}
\caption{ $\gamma$-ray spectrum of {\Olam} (preliminary).
(a) Bound-state region ($-21< -B_\Lambda < 8$ MeV)
is gated. (b) Same as (a) but Doppler-shift correction is applied.
(c) Highly unbound region ($-B_\Lambda >50$ MeV) is gated. 
\label{Ospec}
}
\end{minipage}
\hspace*{2mm}
\begin{minipage}{5.3cm}
\vspace{0.7cm}
\hspace*{-7mm}\includegraphics[width=6.2cm]{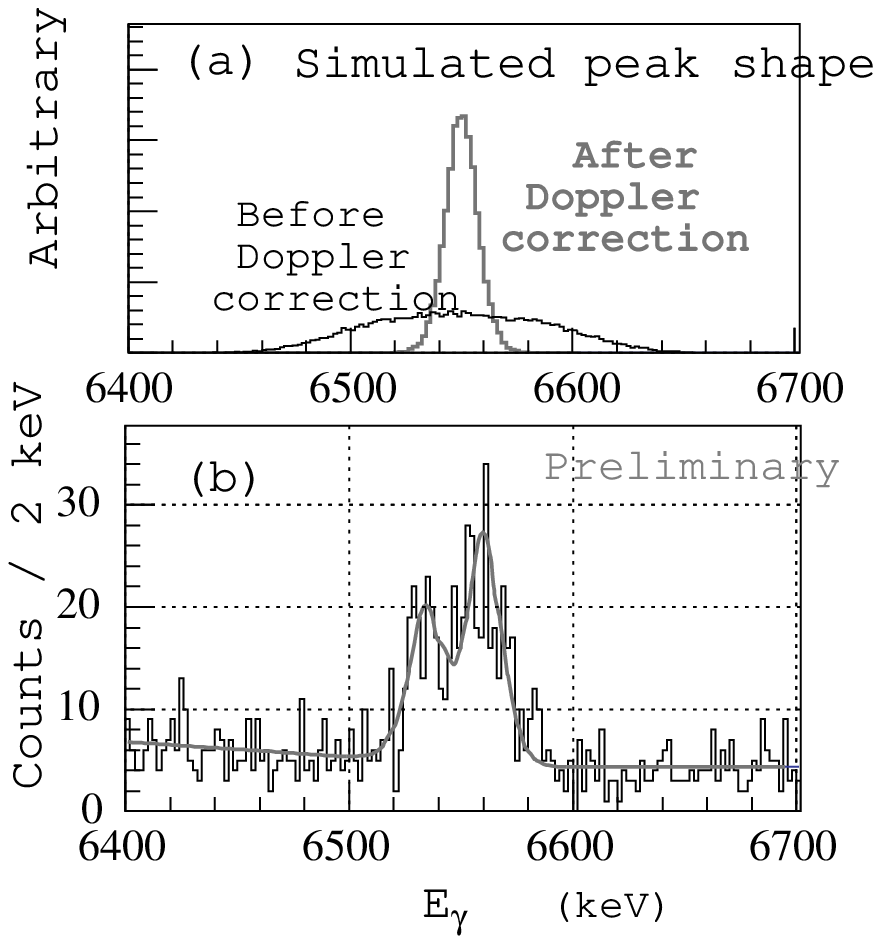}
\vspace*{-8mm}
\caption{
(a) Simulated peak shape for a fast $\gamma$ transition
after Doppler-shift correction.
(b) The structure around 6.55 MeV in 
Fig.~\protect\ref{Ospec} (b)
was fitted with two peaks of the simulated peak shape
(preliminary).
\label{O16FIT}
}
\end{minipage}
\end{figure}

Figure~\ref{Ospec} shows preliminary 
$\gamma$-ray spectra for {\Olam}.
Figure~\ref{Ospec} (a) shows the spectrum when events in the 
6 MeV-excited $1^-$ state region of the {\Olam} mass 
spectrum ($ -21 < -B_\Lambda < 8$ MeV) are selected.
A broad bump is observed at around 6.55 MeV.
After the event-by-event Doppler-shift correction was applied,
the broad bump is resolved into two narrower peaks as shown in (b).
This structure is not observed for the highly unbound region 
($ -B_\Lambda > 50$ MeV) as shown in  Fig.~\ref{Ospec} (c),
in which beam-induced $\gamma$ rays from the $^{16}$O target 
are observed. (The 6130 keV $^{16}$O peak width 
demonstrates the resolution in this energy region.)

The structure at 6.55 MeV is thus
attributed to the $M1$($1^-_2$$\to$$1^-_1,0^-$) 
transitions in {\Olam}.
The peaks in Fig.~\ref{Ospec} (b)
were fitted with the expected Doppler-corrected
peak shape which was calculated from 
a simulation for the Doppler-shift correction.
The spectrum was fitted well with two peaks as shown in Fig.~\ref{O16FIT}.
The energies (and the counts) of these peaks were obtained as
6534.1$\pm$1.5 keV (149$\pm$18 counts) and
6560.2$\pm$1.3 keV (226$\pm$30 counts).
By comparing the ratio of the peak counts
with the expected branching ratios, 
the 6534 keV and 6560 keV peaks were assigned as
$1^-_2$$\to$$1^-_1$ and $1^-_2$$\to$$0^-$ transitions, 
respectively~\cite{UKA03}.
Then we obtained the energy spacing of the ground-state doublet:
$$E(1^-)-E(0^-) = 26.1 \pm 2.0 {\rm ~keV ~(preliminary)}.$$
It is the smallest spacing in hypernuclear fine structure
observed so far.
This very small spacing results
from a cancellation of the spin-spin
force ($\Delta$ term) and the tensor force ($T$ term) contributions.
It gives the tensor term strength of $T=+30$ keV (preliminary)
from Eq.~\ref{Eq:16O} and with the $\Delta$,
$S_\Lambda$, and $S_N$ values already determined
from previous Hyperball experiments.
This is the first experimental information on the $\Lambda$$N$ 
tensor force.

The meson-exchange baryon-baryon interactions models
(ND, NF, NSC89, NSC97f) predict a tensor force strength of  
$T = 18 - 54$ keV
through a G-matrix calculation~\cite{MIL99}.
They are almost consistent with the experimental value.

\section{ Study of {\Btenlam} }

\begin{figure}[t]
\begin{center}
\includegraphics[height=3.8cm]{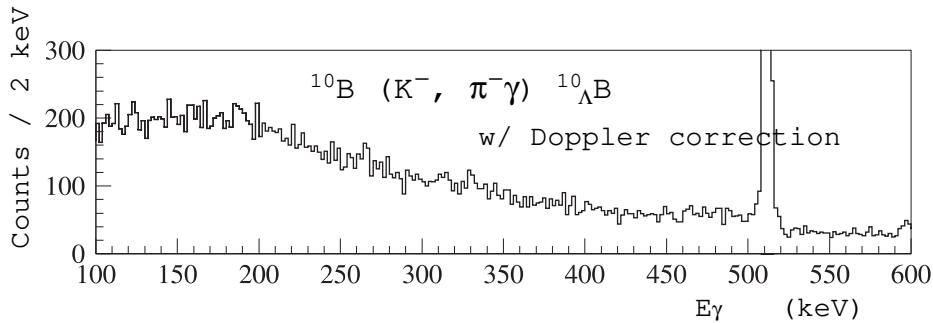}
\vspace*{-8mm}
\caption{ Preliminary $\gamma$-ray spectrum for the bound-state region
of {\Btenlam} ($-23 < -B_\Lambda < 7$ MeV). No peak is observed. 
\label{BTENSPEC}
}
\vspace*{-0.5cm}
\end{center}
\end{figure}

The purpose of the $^{10}$B target run in E930('01)
is to measure the energy spacing of the
{\Btenlam} ground-state doublet ($2^-,1^-$) by observing the
spin-flip $M1$ transition ($2^-$$\to$$1^-$).
Since the production cross section of the $2^-$ state is large enough,
the $M1$ transition can be
easily observed if the level spacing is as large as predicted 
($\sim$200 keV).
On the other hand, if the spacing is smaller than $\sim$100 keV,
the $\gamma$ transition is overcome by weak decay.

Figure \ref{BTENSPEC} is a preliminary $\gamma$-ray
spectrum when the bound-state region of 
$^{10}$B($K^-,\pi^-$){\Btenlam}
is selected.
We observed no peak structure.
Considering the number of the expected $\gamma$-ray peak yield,
we concluded that the $2^-$ state is higher than the $1^-$ state
only by 100 keV or less, or the order of the spins in the doublet is
reversed.
Thus, we confirmed the old result by Chrien {\it et al.}~\cite{CHR90}
with higher statistics.
The confirmed result of
$E(2^-) - E(1^-) < 100$ keV seems contradictory to the
$\Delta$ value (0.4 MeV) obtained from the
{\Lilam} ground-state doublet (3/2$^+$,1/2$^+$) spacing,
suggesting that more theoretical and experimental studies are necessary,
particularly for the $\Sigma$-$\Lambda$ coupling
effect as investigated in Ref.~\cite{MIL04}.

\section{ Complete study of {\Lilam}}

\begin{figure}[t]
\begin{center}
\includegraphics[height=7.4cm]{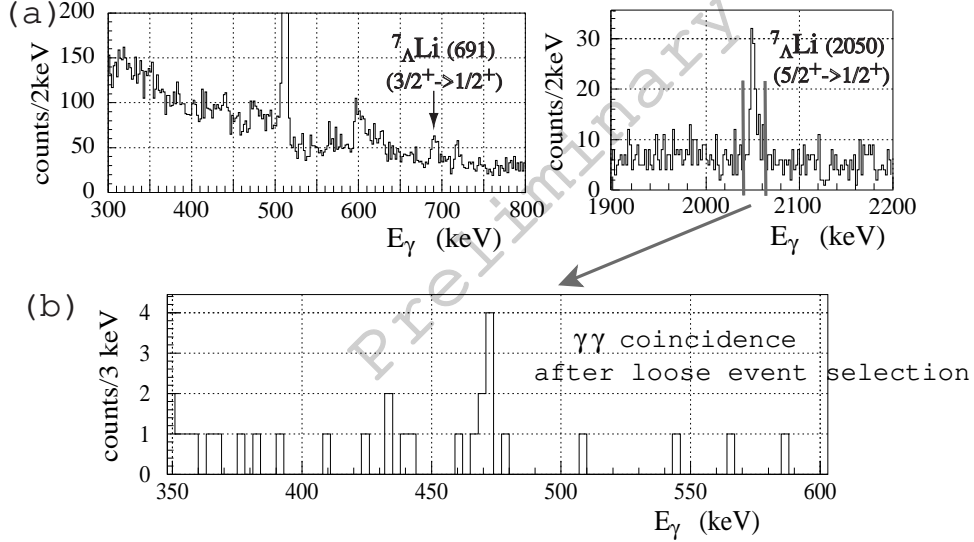}
\vspace*{-0.5cm}
\caption{(a) $\gamma$-ray spectrum for the unbound region 
($0 < -B_\Lambda <-40$ MeV) of {\Btenlam}. 
Two $\gamma$-ray peaks from {\Lilam}
produced as hyperfragments are observed.
(b) $\gamma$-ray spectrum in coincidence with the {\Lilam} 
$E2$($5/2^+ \to 1/2^+$) $\gamma$-ray peak.
\label{GAMMAGAMMA}
}
\end{center}
\end{figure}

In the $^{10}$B target data in E930('01),
we observed $\gamma$ rays from {\Lilam} produced as hyperfragments
from highly excited states of {\Btenlam}, presumably though
the $s$-substitutional $^{10}$B ($3^+$, $\sim$28 MeV excited)
state decaying into {\Lilam} + $^3$He as shown in Fig.~\ref{ALLLEVEL} 
(d). Figure~\ref{GAMMAGAMMA} (a) shows the $\gamma$-ray spectrum
when the unbound region ($0 < -B_\Lambda < 40$ MeV) was selected.
The $M1$($3/2^+ \to 1/2^+$) and $E2$($5/2^+ \to 1/2^+$)
$\gamma$ rays of {\Lilam} previously observed in E419 are identified.
We selected the $E2$ $\gamma$-ray events 
($2042 < E_\gamma < 2058$ keV)
and plotted a spectrum of another $\gamma$ ray
emitted in coincidence.
As shown in Fig.~\ref{GAMMAGAMMA} (b),
a peak was observed at 471 keV.
The probability that background fluctuation makes such a peak 
anywhere in the region of 0.1--1 MeV
is 0.006\%. This peak is assigned as the $M1$(7/2$^+$$\to$5/2$^+$)
transition, because it is the only transition
emitted in coincidence with $E2$(5/2$^+$$\to$1/2$^+$).
This is the first successful
application of the $\gamma$-$\gamma$ coincidence method
to hypernuclei.

\begin{figure}[h]
\begin{center}
\includegraphics[height=6.65cm]{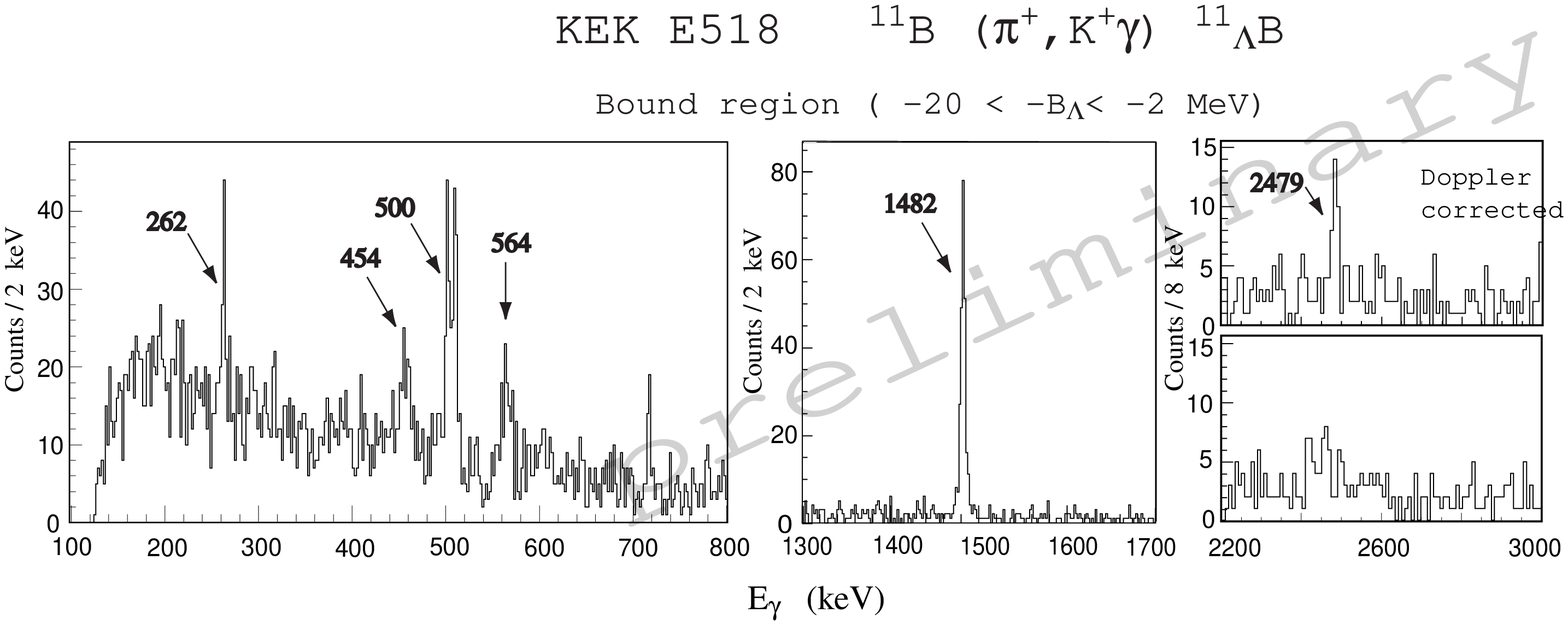}
\vspace*{-5mm}
\caption{ Preliminary
$\gamma$ ray spectra of $^{11}$B$(\pi^+, K^+ \gamma)${\Blam}
experiment at KEK (E518).
Six $\gamma$ rays of {\Blam} are observed.
\label{B11SPEC} 
}
\end{center}
\end{figure}

The observed energy can be compared with theoretical predictions.
A cluster-model calculation by Hiyama {\it et al.}
predicted the (7/2$^+$,5/2$^+$) spacing to be 560 keV 
when the $\Lambda$-spin-dependent 
spin-orbit force is assumed to be
zero \cite{HIYHYP00}. It is close to the observed value.
According to the Millener's shell-model calculation \cite{MIL04},
the energy spacing is described as 
$E(7/2^+) - E(5/2^+) = 1.29\Delta + 2.20 S_\Lambda + 0.02 S_N - 2.39 T
+ \Lambda\Sigma$. 
By using the already determined 
values of $\Delta$, $S_\Lambda$, $S_N$, $T$, and the theoretically
calculated $\Lambda\Sigma$ 
value,
the equation gives 511 keV, being also close to the observation.
It is found that the observed value 
is consistent with the already-known strengths of the
spin-spin force ($\Delta$) and the 
very small spin-orbit force ($S_\Lambda$).

Together with the  E419 results \cite{TAM00},
we have clarified the complete level scheme
and energies of all the bound states of {\Lilam}
as shown in Fig.~\ref{ALLLEVEL} (a).

\section{Spectroscopy of {\Blam} (E518)}

In 2002,
we carried out a $\gamma$ spectroscopy experiment of $^{11}_\Lambda$B 
with the $(\pi^+,K^+)$ reaction at 1.05 GeV/$c$
employing Hyperball and the SKS spectrometer at KEK-PS \cite{MIU03}.
One of the purposes of this experiment is to 
measure the transition probability $B(M1)$ of the $\Lambda$ spin-flip
$M1$ transition $^{11}_\Lambda$B($3/2^+ \to 1/2^+$) 
and extract information on the 
magnetic moment of a $\Lambda$ inside a nucleus by the method
described in Ref.~\cite{TAM01}.
The other purpose is to cross-check
the $\Lambda$$N$ spin-dependent interaction parameters
which have been determined from the {\Lilam}, {\Belam}, and {\Olam} 
experiments with Hyperball.

The experimental setup is almost identical to the one in
E419~\cite{TAM00,TAN01}. 
We used a 10 cm-thick 98\%-enriched
$^{11}$B metal target.
When the bound-state region is gated in the {\Blam} mass spectrum,
the $\gamma$-ray spectrum exhibited six peaks
as shown in Fig.~\ref{B11SPEC}.
One of them was observed in the Doppler-shift-corrected spectrum.
They are attributed to transitions from $^{11}_\Lambda$B, but
the assignment of all these $\gamma$ rays and
the reconstruction of the level scheme are difficult
because of low statistics which does not allow
$\gamma$$\gamma$ coincidence
measurement.

The prominent peak at 1482 keV
is assigned as $E2(1/2^+ \to 5/2^+)$
(see Fig.~\ref{ALLLEVEL} (e)).
It is likely an $E2$ transition because its narrow width indicates a
lifetime of the transition longer than $\sim$10 ps,
which gives a very small $B(M1)$ value if it is an $M1$ transition.
The $1/2^+ \to 5/2^+$ transition is the only $E2$ transition
expected in {\Blam}, and the observed largest $\gamma$-ray yield 
is also consistent with this assignment.
It is to be noted that
the shell-model prediction by Millener~\cite{MIL04}
for this $E2$ energy with the experimentally determined
$\Lambda$$N$ interaction parameters is 1020 keV,
significantly lower than the observed energy.

\section{Hyperfragments (E509)}

In KEK-PS E509, 
we attempted an experiment of 
inclusive $\gamma$-ray measurement
in the stopped $K^-$ absorption reaction, which is known to produce
various hyperfragments with large production yields.
See Ref.~\cite{TAN03} for details.
We stopped $K^-$ from the K5 beam line 
on several light targets ($^7$Li, $^9$Be, $^{10}$B, $^{11}$B,
and $^{12}$C)
and measured $\gamma$ rays with Hyperball.
From the $^{10}$B, $^{11}$B, and $^{12}$C targets,
we observed the $^7_\Lambda$Li($5/2^+\to1/2^+$) 
transition at 2050 keV.
The yield of this $\gamma$ ray for the $^{10}$B target
is very large, 500 counts in 3.5 days' beam time,
suggesting the effectiveness of this method.
The production rate of the {\Lilam}($5/2^+$) state is
derived to be 0.075$\pm$0.016\% per stopped {\km} on $^{10}$B target.

This method is powerful
with a larger Ge detector array,
by which the $\gamma \gamma$ coincidence method allows
detection of hypernuclear $\gamma$ rays and their assignments.  
It may open a new possibility to study various hypernuclei
including neutron/proton-rich ones which cannot be
produced by the direct reactions such as {\kmpim}
and {\pipkp}.

\begin{figure}[t]
\begin{minipage}{80mm}
\includegraphics[width=7.8cm]{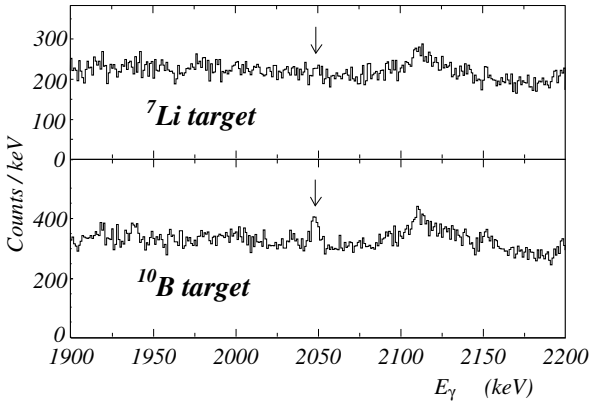}
\end{minipage}
\hspace*{4mm}
\begin{minipage}{71mm}
\vspace*{-5mm}
\caption{$\gamma$ ray spectra in the ($K^-_{stop}, \gamma$)
reaction on $^7$Li and $^{10}$B targets.
The 2050 keV $\gamma$ ray from {\Lilam} is 
abundantly observed in the $^{10}$B target spectrum.
\label{E509}
}
\end{minipage}
\end{figure}

\section{Future Plans}

We are now constructing Hyperball2, an upgraded version of Hyperball
for near-future experiments at KEK and BNL.
It has an efficiency twice as large as the present Hyperball,
realizing a $\gamma$$\gamma$ coincidence efficiency 
larger by four times. It will play an essential role in study of
hypernuclei having complicated level schemes such as {\Blam}.

At the 50 GeV proton accelerator facility at J-PARC,
we plan to pursue various types of
hypernuclear $\gamma$  spectroscopy experiments~\cite{LOI,NAG04}. 
We have started development of
faster readout techniques and faster background-suppression counters
necessary for the stronger beams expected at J-PARC.

\section{Summary}

We have investigated various $p$-shell $\Lambda$ hypernuclei
employing Hyperball.
In E930('01), we successfully observed two $\gamma$ transitions
of $1_2^- \to 1_1^-,0^-$ at 6.55 MeV, and the ground-state doublet 
$(1_1^-,0^-)$ spacing was obtained to be 
$E(1_1^-) - E(0^-)$ = 26.1$\pm$2.0 keV.
It gives the $\Lambda$$N$ tensor force strength of $T = $ +30 keV.
All the $\Lambda$$N$ spin-dependent force parameters have been
thus determined from our $\gamma$ spectroscopy experiments.
In $^{10}$B($K^-,\pi^-\gamma$) data, we observed 
$\gamma$ rays from hyperfragments
such as {\Belam}($3/2^+ \to 1/2^+$) and {\Lilam}($7/2^+ \to 5/2^+$).
In the observation of the {\Lilam}($7/2^+ \to 5/2^+$) transition,
we successfully applied the $\gamma\gamma$ coincidence method 
to hypernuclei for the first time.
On the other hand, the spin-flip M1 transition of
{\Btenlam}($2^- \to 1^-$) was not observed.
At KEK, we studied {\Blam} with the {\pipkp}
reaction and observed six $\gamma$ transitions.
We also performed a pioneering experiment 
with the ($K^-_{stop}$, $\gamma$) reaction (E509),
and observed a $\gamma$-ray peak from {\Lilam} hyperfragments.
Hypernuclear $\gamma$ spectroscopy will be further pursued 
with much stronger beams at J-PARC.

\end{document}